# Chemical leaching of Al-Cu-Co decagonal quasicrystals

Shashank Shekhar Mishra[1] and Thakur Prasad Yadav[1,2]

[1]Hydrogen Energy Centre, Department of Physics, Banaras Hindu University, Varanasi-221005, India

[2]Department of Physics, Faculty of Science, University of Allahabad, Prayagraj-211002, India

**Abstract**

In the present investigation, the chemical leaching of the poly-grain $Al_{65}Cu_{15}Co_{20}$ and $Al_{65}Cu_{20}Co_{15}$ decagonal quasicrystalline alloy have been studied. The polished surfaces of as-cast alloys were leached with 10 mole NaOH solution for 0.5- 8 hours. The x-ray diffraction, scanning electron microscopy and transmission electron microscopy techniques have been used for structural and microstructural characterization. Energy dispersive x-ray analysis has been carried out for chemical composition analysis. Chemical leaching exclusively removes the Al from the surfaces of the $Al_{65}Cu_{15}Co_{20}$ and $Al_{65}Cu_{20}Co_{15}$ decagonal quasicrystalline alloys, consequently by the formation of porous structure containing nano size particles of Cu, Co and their oxides have been observed. $Al_{65}Cu_{15}Co_{20}$ exhibits high porosity in comparison to $Al_{65}Cu_{20}Co_{15}$ alloy, however the size of the precipitated nano-particles *i.e.* Cu, Co, $Cu_2O$ and CuO were smaller in the case of $Al_{65}Cu_{20}Co_{15}$ alloy. The energy dispersive X-ray analysis mapping suggests homogeneous distribution of Cu and Co found on the leached surface and the presence of oxygen was also detected.

**Key words**: -quasicrystal, decagonal phase, leaching, Al-Cu-Co, dealloying, nano-particles.

## 1. Introduction

The quasicrystalline (QC) materials have many interesting and unique properties such as high hardness [1, 2], low surface energy [3], low electrical and thermal conductivity [4, 5], low friction coefficient [6, 7]. These unique properties are helpful to use the QCs in different relevant applications [8, 9]. Participation of catalytic elements (such as Fe, Cu, Co, Cr, Ni, Pd etc. in many Al-based QCs), along with stability at high temperature and high brittleness [1-10], furnish QC materials have additional features to use as a catalyst [11, 12]. However, the bulk Al-based QCs alloys have poor catalytic activity, as surface has been covered by passive aluminum oxide layer [13]. Moreover, in several reports the carbon nano-materials have been synthesized using QC as catalyst [14-16]. The QC materials have revealed a better catalytic activity after activation and this activation was achieved by chemical leaching with alkaline solution (NaOH), which is a captivating process leaching the nano particles with ultra-fine pores [17]. In a single-phase Al-Cu-Co decagonal alloy / intermetallic compound precursor, the porous Cu-Co was produced by chemical leaching process [17]. Due to their structural and functional properties QC supported porous metals act as improved catalyst [18].

The catalytic behavior of Al-based QC materials for steam reforming of methanol (SRM) has been patented, for a large amount of hydrogen production using leached $Al_{65}Cu_{20}Co_{15}$ and $Al_{65}Cu_{20}Fe_{15}$ alloys [19]. A better catalytic activity of Al-based QCs was observed in Al-Cu-Fe (-

Co/Cr) in comparison to Cu/ZnO/$Al_2O_3$ in the temperature range of 548K – 598K [20]. The catalytic activity of other Al-based QC *i.e.* Al-Pd-Mn has been examined for the methanol decomposition reaction and the rate of hydrogen production was higher with QC in comparison to crystalline version [19]. Several others studies were also shown the better catalytic properties of QC alloys after leaching [21-23]. The Ti-based QC *i.e.* $Ti_{45}Zr_{35}Ni_{17}Cu_3$ also acts as a better catalyst for the oxidation of cyclohexane [24]. The Al deficient $Al_{63}Cu_{25}Fe_{12}$ QC alloy after leaching with alkaline solution reveals superior catalytic activity for SRM compared to the industrial Cu-based catalyst [22, 25]. Further, it was reported that the catalytic performance of $Al_{63}Cu_{25}Fe_{12}$ alloy leached with $Na_2CO_3$ was better as compared to same alloy leached with NaOH at 663K, however at lower temperature (513K) the result was opposite [26]. The catalytic action of $Al_{63}Cu_{25}Fe_{12}$ alloy was enhanced by leaching of wet milled QC powder [27] and it was also very stable, in fact better than known ZnO/Cu catalyst [28]. The cause of high performance was the formation of leached layer consisting of Cu, Fe and their oxides. The Fe species on the leached layer offers the stability and enhances the catalytic activity this also prevents the agglomeration of Cu on the surface [28]. Residual Al is also a key factor for high catalytic performance of $Al_{63}Cu_{25}Fe_{12}$ QC alloy [28]. Calcinations at 773K stimulates the creation of fine Cu nano-particles on the QC surface exposed to leaching treatment which in turn amplifies the catalytic activity about five times [29]. It has been observed that the five-fold axes of Al-Cu-Fe system offers low resistance against leaching and when this surface was leached for a longer time, the surface become affluent with $Fe_3O_4$ [30]. The study of leaching effect shows that in case of Al-Ni-Co decagonal QC, Al preferentially removes along the high symmetry ten-fold axis than the two-fold axis [31].

Up to now the exact chemical mechanism liable for catalytic properties of QC materials is not well defined. It has been assumed that the formation of fine Cu and Fe particles due to removal of Al by leaching treatment from the surface of QC powder, is the key factor for catalytic activity [28]. The agglomeration of Cu particles is barely observed at the leached QC surface due to immiscibility of Cu with Fe (or Cu with Co), as a result, a plenty of Cu/Fe sites at the surface get observed after leaching [32]. As compared to the QC, the corresponding approximants have been found to show lower catalytic activity. One example is Al-Cu-Co where ternary QC systems was leached and most of the Al is eliminated from the lattice sites at the surface leaving behind a nicely dispersed structure of the rest of the elements such as Cu and Co in contest to the binary system (Cu-Co), the elements are very less soluble so it is tough to get the Cu-Co binary solid solution by conventional process [33].

In the present investigation, the effect of the chemical leaching by the alkaline solution (10 mole NaOH) on two decagonal QC (DQC) alloys namely $Al_{65}Cu_{15}Co_{20}$ and $Al_{65}Cu_{20}Co_{15}$ have been studied in detail. The different structural evolution as a function of leaching time of the two alloy compositions will be described and discussed.

## 2. Experimental procedures

The alloys with nominal composition $Al_{65}Cu_{15}Co_{20}$ and $Al_{65}Cu_{20}Co_{15}$ were prepared from highly pure (99.99 %) metals (Al, Cu and Co) which were melted in conventional radio frequency induction furnace under argon atmosphere. The alloys were re-melted three times to assure the homogeneity. The surfaces were polished before each leaching treatment using diamond paste of successively smaller size of 6 to 0.25 μm. Leaching was performed in ambient conditions by placing droplets of NaOH aqueous solution with 10 mole concentration on the surfaces using a pipette. Leaching was performed at various times from 30 min to 8 hours. The leached samples were washed with double distilled water and methanol in ultrasonic bath until no alkali nature was detected. The structural characterization of unleached and leached alloys were carried out by x-ray diffraction (XRD) technique where Philips 1710 XRD system was used with CuKα (λ=1.540Å) radiation. The surface morphology of unleached and leached alloys were examined by scanning electron microscopy (SEM Quanta 200, operating at 25kV). The details microsture along with phase analysis of all the alloys were carried out using transmission electron microscope (TEM TECNAI 20G$^2$) operated at 200kV in imaging and diffraction mode.

## 3. Results and Discussion

The phase formation in the as-cast alloy as well as of the leached surface of the sample have been examined by XRD and TEM diffraction methods. The XRD of the as-cast $Al_{65}Cu_{15}Co_{20}$ sample showed mostly sharp diffraction peaks all the peaks could be indexed as the DQC using Mukhopadhyay and Lord indexing model as shown in figure 1(a) [34]. In the present case, the quasilattice ($a_R$) and the periodic lattice (c) constants were also calculated and have been found to be 3.98 and 8.15 Å respectively. Here, no other crystalline phases have been observed, which indicated that a high quality of single phase DQC can be produced from the $Al_{65}Cu_{15}Co_{20}$ alloy by simple melting. Figure 1(b-f) shows the XRD patterns of the leached DQC alloys with 10 mole NaOH aqueous solutions for different time, it is evident that leaching with the NaOH solution results in the broadening of the peak as well as appearance of some new peaks corresponding to Co and Cu, in the DQC phase, however the peaks remained sharp even in the 8 h leached alloy. No visible peaks from copper oxides/ cobalt oxides were observed after leaching treatment. It become clear that alkali leaching considerably reduces Al content from the decagonal surface, the following chemical reaction of NaOH with Al will take place on the surface;

$$2Al + 2NaOH + 2H_2O = 2NaAlO_2 + 3H_2 \qquad (i)$$

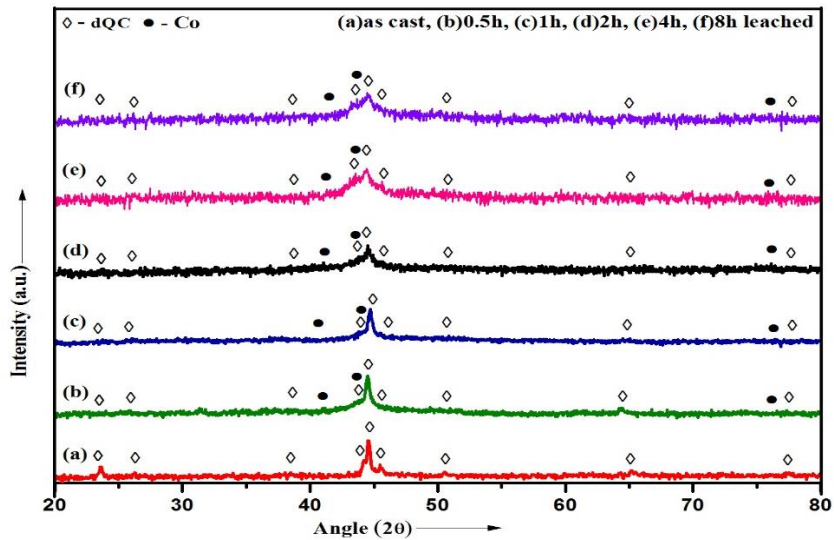

*Fig.1 XRD pattern of $Al_{65}Cu_{15}Co_{20}$ alloy (a) as-cast (b) 30min (c) 1h (d) 2h (e) 4h (f) 8h leached.*

Therefore, the alloy surface will be Al deficient and reconstructed into deformed DQC phase. The precipitation of Co and Cu metallic phases has been observed on the DQC surface as detected in the XRD patterns shown in figure 1(b-f). However, the peak corresponding to Co is more prominent as compare to Cu. Although the DQC XRD peak in figure 1(a-f) are similar but there are some differences between the peak position and peak broadening for the above said two phases. Therefore, the local magnifications of figure 1 of the strong reflection peak (102202) plane of the DQC phase have been shown in figure 2.

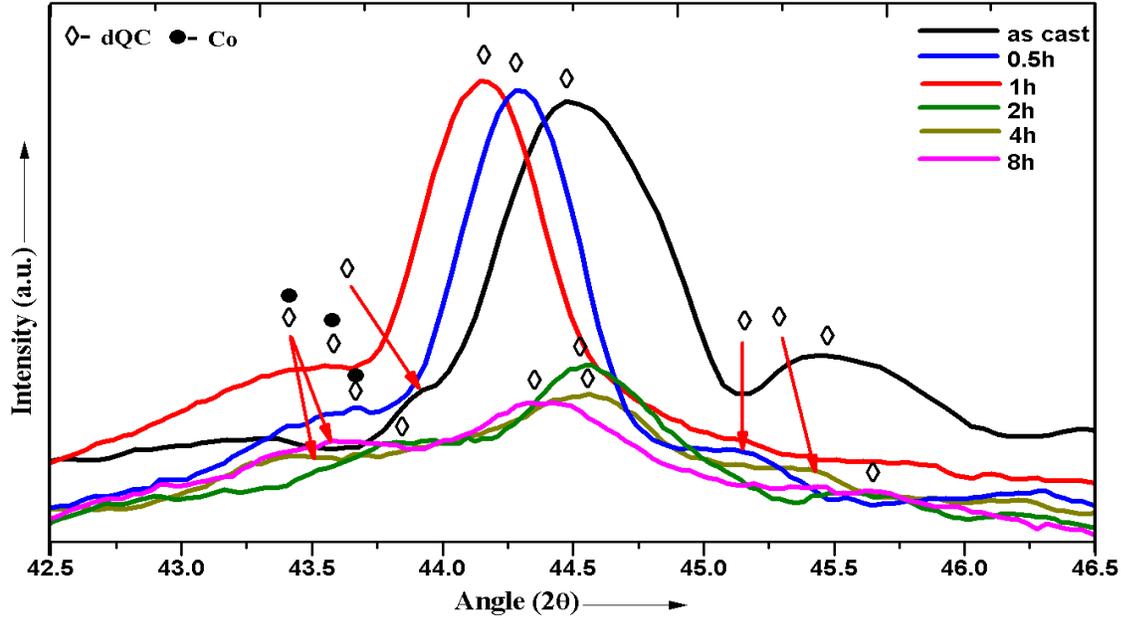

*Fig.2 Magnified view of XRD pattern of $Al_{65}Cu_{15}Co_{20}$ alloys.*

The peak intensity of the DQC phase is nearly constant upto 1 h of leaching and then decreased with the increase leaching time. However, the peak position is sifted towards lower angle side up to 1h leaching Moreover, the peak position of the 2-8 h leached sample is very closed to the as-cast alloy. It indicated that the structural modifications of the DQC exist in the leached samples. As is expected clear, during leaching treatment the Al will be dissolved and surface of the alloys having Al deficient deformed DQC phase and further after few layers of complete leaching the constituent elements (Co & Cu) appeared on the surface along with ordered DQC under layer. Such type of deformation in Al-Cu-Co, DQC phase in the direction of ten-fold symmetry axis is also possible by low temperature annealing [35]. The leaching treatment is associated with the presence of antiphase domain boundaries inside the DQC matrix in form of higher concentrations of Co & Cu. The antiphase domain boundaries make the initiation of dislocation movement easier and an increase in the density of dislocations increases the deformation area in the DQC alloy.

Figure 3(a-f) shows the XRD patterns of as-cast Cu rich decagonal alloy *i.e.* $Al_{65}Cu_{20}Co_{15}$ and chemically leached with 10 mole NaOH aqueous solutions for different time ( 0.5, 1, 2, 4 and 8 h ) alloys respectively. All the diffraction peaks of the as-cast alloy have been indexed by DQC phase according to their inter-planar spacing (*d*- spacing) and relative intensity [34]. No other crystalline phases have been observed. It indicates that the DQC phase is formed directly from the $Al_{65}Cu_{20}Co_{15}$ alloy composition also under the present processing condition.

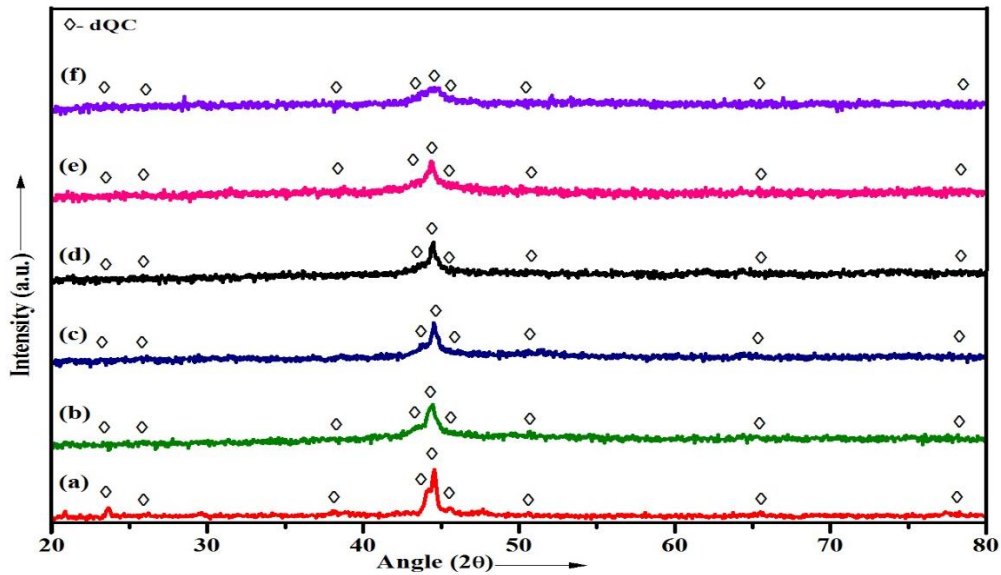

*Fig.3 XRD pattern of $Al_{65}Cu_{20}Co_{15}$ alloy (a) as-cast (b) 30min (c) 1h (d) 2h (e) 4h (f) 8h leached.*

Figure 3(b-f) shows powder XRD patterns of the DQC alloys subjected to different hours of leaching treatments; here no significant changes have been observed peaks and all diffraction peaks corresponded mainly to a single DQC phase. There are no separate peaks associated with Cu and Co or their oxides, implying that these are only present in highly disordered states, therefore diffraction peaks appear to become invisible due to line-broadening. On the other hand, for the leached DQC, $Al_{65}Cu_{15}Co_{20}$ alloy, peaks due to Co and Cu phases appear as shown in figure 1. However, a low angle shifting has been observed in XRD peak (around $2\theta = 44.5°$) as shown in figure 4 (the close view of the maximum intense DQC peak after 0.5 h of leaching). The peak intensity of all the leached alloys has been reduced significantly. The XRD peak of leached alloys for longer leaching times *i.e.* 1, 2, 4 and 8h also shows a small shifting towards lower angle side as shown in figure 4. Here again the alloy surface will be Al deficient after leaching and reconstructed into deformed DQC phase like $Al_{65}Cu_{15}Co_{20}$. However, in the Cu rich decagonal alloy *i.e.* $Al_{65}Cu_{20}Co_{15}$ the leaching kinetics was fast.

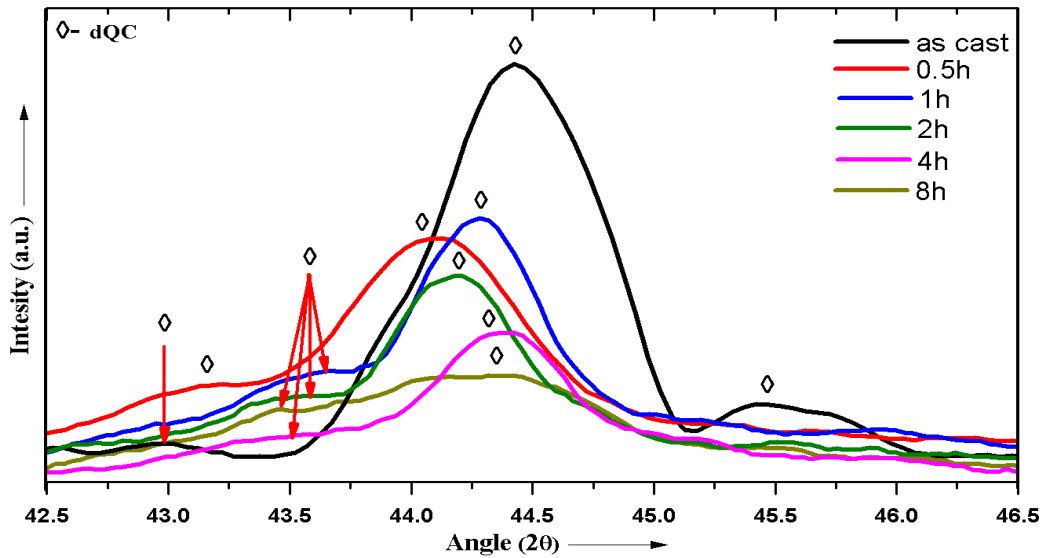

*Fig.4. Magnified view of XRD pattern of $Al_{65}Cu_{20}Co_{15}$ alloys.*

The broad XRD peaks have been observed from all of the leached DQC alloys as shown in Figure 1 and Figure 2. These peaks were identified as belonging to Co, Cu and DQC phase. The crystallographic structures of all leached surface have changed from DQC phase to Co, Cu and deformed DQC phase. Therefore, the leaching process can be considered as a sort of the formation of nano crystallite on the DQC surface that occurs due to the compositional change as outlined in equation (i). The broadening of the XRD reflections after leaching clearly indicated that surface of the DQC alloys containing nano crystallite or highly strained. Therefore, the crystallite size and lattice strain have been calculated from the broadening of the XRD reflections using single line profile analysis [36]. A sharp reduction in the crystallite size with leaching time has been noticed for $Al_{65}Cu_{20}Co_{15}$ alloy. However, in the $Al_{65}Cu_{15}Co_{20}$ alloy composition the reduction in the crystalline size was sluggish. The initial leaching time, lattice strain gets decreased and become minimum for 2h leaching which further increase with leaching time. The lattice strain for both the leached alloys *i.e.* $Al_{65}Cu_{20}Co_{15}$ & $Al_{65}Cu_{15}Co_{20}$ was fairly low. Therefore, the broadening in XRD peak is mainly due to precipitation of the nano crystallite on the DQC surface.

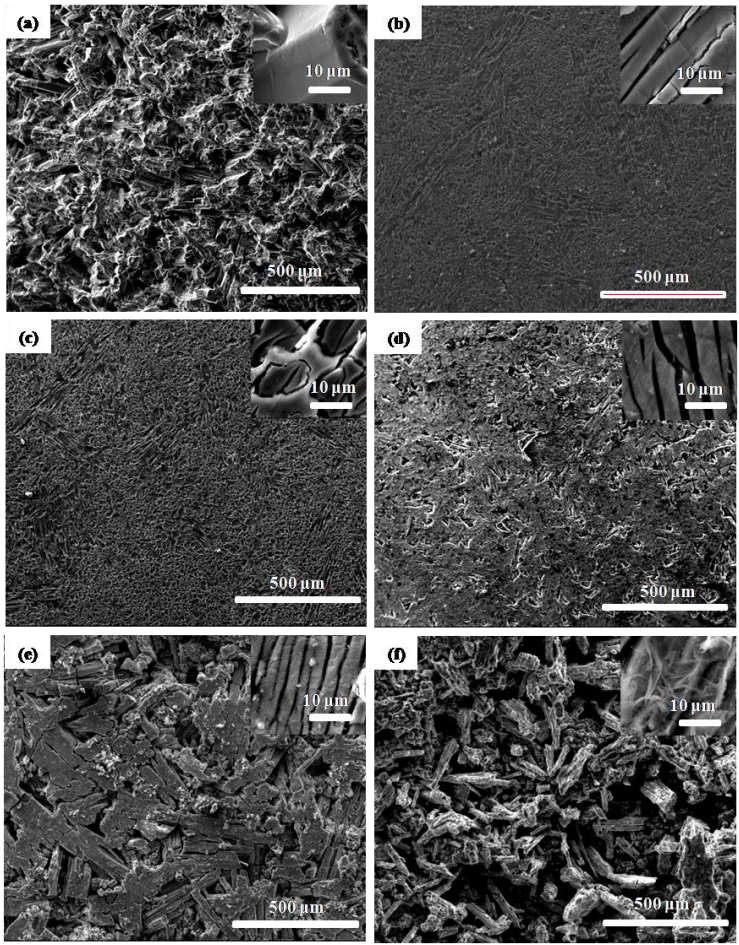

*Fig.5 SEM micrograph of $Al_{65}Cu_{15}Co_{20}$ (a) as -cast (b) 30 min (c) 1h (d) 2h (e) 4h (f) 8h leached with 10 mole NaOH solution.*

The detailed surface microstructural characterization of $Al_{65}Cu_{15}Co_{20}$ as-cast as well as leached DQC alloys have been done by SEM. Figure 5(a-f) shows the surface morphology of the as-cast and leached surface the leaching has been done by employing 10 mole NaOH alkaline solution for 0.5, 1, 2, 4 and 8 h respectively. The magnified view of the microstructure has been given in the inset of the corresponding SEM image. Figure 5 (a) shows the SEM micrograph of as-cast polished surface and corresponding inset is the magnified view of the fractured surface. A typical decagonal prismatic morphology has been observed 50 µm in diameter and 500 µm in length where the growth direction was found to lie along the 10-fold direction. SEM observation of the alloy after leaching at different time (figure 5(b-f)) suggests that it occurred homogeneously on the DQC surface, outlining the columnar structure and resulting in the formation of a sub-nanometer porous/particle on the surface compared to the as-cast alloy. The stronger the anisotropy of the leaching kinetics of the two different orientated surfaces *i.e.* 2-fold and 10-fold of the polygrain DQC leads to fast leaching in the periodic direction along the needle axis. However, for longer leaching time *i.e.* 8h the quasiperiodic direction also gets leached out.

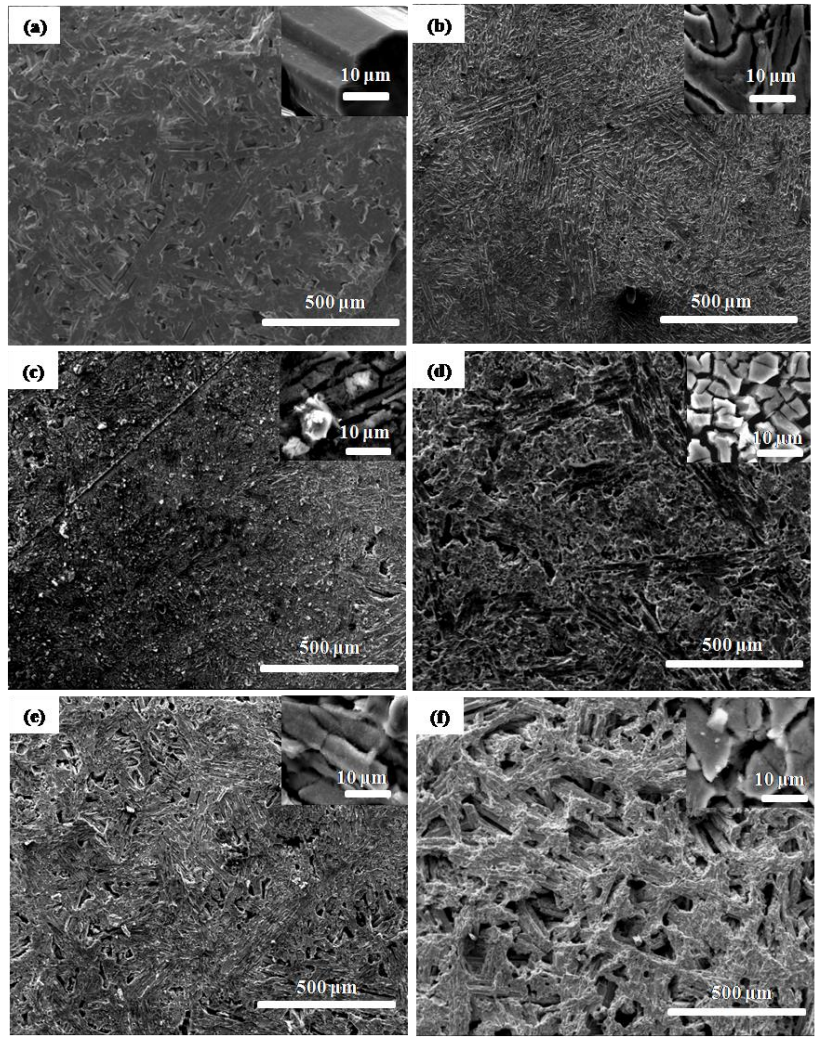

*Fig.6 SEM micrograph of $Al_{65}Cu_{20}Co_{15}$(a) as-cast (b) 30 min (c) 1h (d) 2h (e) 4h (f) 8h leached with 10 mole NaOH solution.*

It appears that 2-fold surfaces are leached efficiently than the quasiperiodic 10-fold surface. Therefore, significant removal of Al from the 2-fold surfaces is taking place and the evolution of Co and Cu particles on the surface become visible. A similar type of observation has been also reported on Al-Co-Ni DQC alloys where leached microstructure indicates preferential leaching along the periodic directions and crystallographic direction rather than macroscopic surface orientation is the key influence on leaching kinetics of quasicrystals [31]. In order to check whether the columnar microstructure on the leached DQC was induced by surface composition or phase, we studied the influence of leaching on microstructure of polygrain Cu rich $Al_{65}Cu_{20}Co_{15}$ DQC alloy. Figure 6 (a-f) shows the SEM microstructures of DQC $Al_{65}Cu_{20}Co_{15}$ as-cast and leached alloys surface where leaching was done with 10 mole NaOH alkaline solution for 0.5, 1, 2, 4 and 8 h respectively. The leached surface did not show many differences, which may suggest that shape of columnar microstructure is not induced by surface composition but by crystallographic directions. However, the leaching kinetics of the samples is somewhat different, which suggests Al is preferentially removed at low leaching time and enhanced at grain boundaries. Therefore, just after 30 min of leaching, the fine particles of Co and Cu content is evolved on the surface as could be clearly seen in figure 6(b-f). A high magnification SEM image within a single grain has been shown in the inset of each microstructure (figure 6(b-f)) demonstrating nanoparticles on the surface.

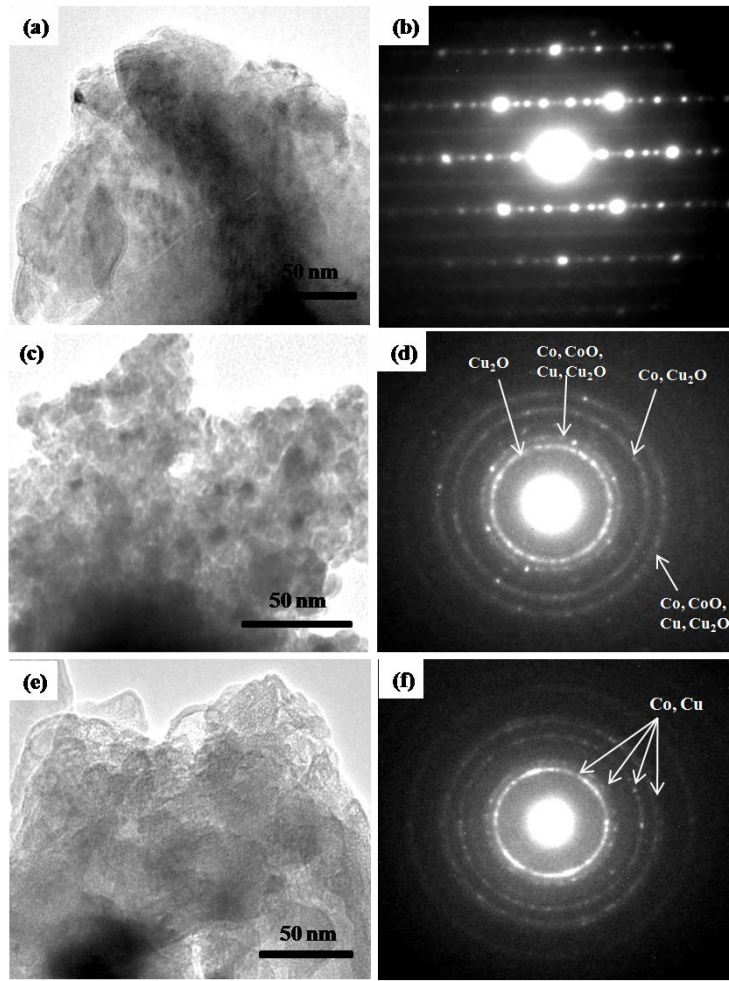

*Fig.7 TEM micrograph of $Al_{65}Cu_{15}Co_{20}$ (a) as-cast (c)2h and (e) 8h, leached with 10 mole NaOH and corresponding diffraction pattern.*

The microstructures and phases of as-cast and leached $Al_{65}Cu_{15}Co_{20}$ and $Al_{65}Cu_{20}Co_{15}$ DQC alloys were also monitored through extensive transmission electron microscopic investigations. Figure 7 (a) show microstructures of as-cast $Al_{65}Cu_{15}Co_{20}$ alloy. Corresponding selected area electron diffraction (SAED) patterns is shown in figure 7(b). The D-type two-fold SAED patterns of the decagonal phase have been observed from several areas of the alloy. The presence of streaking along the aperiodic direction, as typified by the streaked diffraction rows was also observed in each SAED as shown in figure 7(b). It may be mentioned that there are two types of two-fold patterns perpendicular to ten-fold zone axis of decagonal phase, namely P-type and D-type in the notation [37]. Figure 7(c &e) shows the microstructures of the leached $Al_{65}Cu_{15}Co_{20}$ DQC alloys leaching for 2 and 8 h respectively and corresponding **to** SAED patterns have been shown in figure 7(d & f). It is clear from the SAED patterns that the leached area produces rings, instead of spots as shown in figure 7(d &f), the ring patterns suggest that the leaching treatment yields nano-grain microstructures aligned randomly in the leached area. The 5 to 20 nm size particles have been observed in the leached microstructure and corresponding SAED patterns have been indexed with Cu, Co, $Cu_2O$ and CoO crystalline system. This indicates that Al gets dissolved from the leached area and the precipitation of nano particles of Cu, Co and their oxide takes place in the leached area.

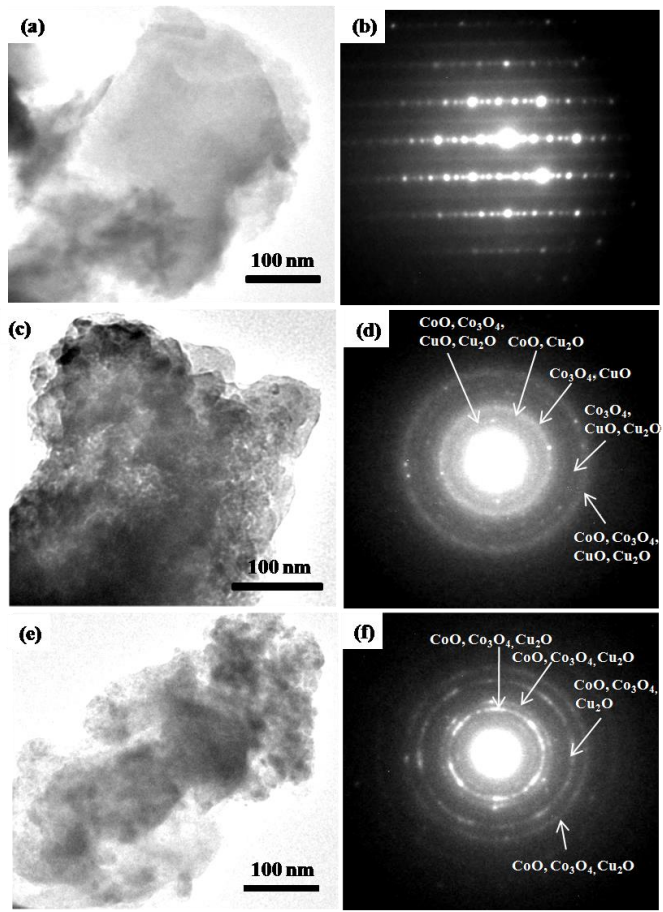

*Fig.8 TEM micrograph of $Al_{65}Cu_{20}Co_{15}$(a) as-cast (c)2h and (e) 8h, leached with 10 mole NaOH and corresponding diffraction pattern.*

Figure 8 (a) shows microstructures of the as-cast $Al_{65}Cu_{20}Co_{15}$ DQC alloy and corresponding SAED patterns in figure 8(b). In this alloy composition also the D-type SAED patterns show the presence of streaking/diffuse row perpendicular to the periodic direction. It is interesting to note that as we increase the Cu content in the DQC Al-Cu-Co alloys system, the streaking gets sharper and intense. The streaking can be explained in terms of the substitutinal disordering of Cu/Co in quasi-periodic planes. The microstructures of the leached $Al_{65}Cu_{20}Co_{15}$ DQC alloys leaching for 2 and 8 h have been shown in figure 8(c &e) and corresponding SAED patterns in figure 8(d & f) respectively. Here also the formation nano particles have been noticed in the leached area and corresponding SAED shows the presence of Cu, Co and their oxides. The tendency of the formation of oxide phase in this alloys composition after leaching with same experimental conditions was higher. However, size of the precipitated particle was small in this case.

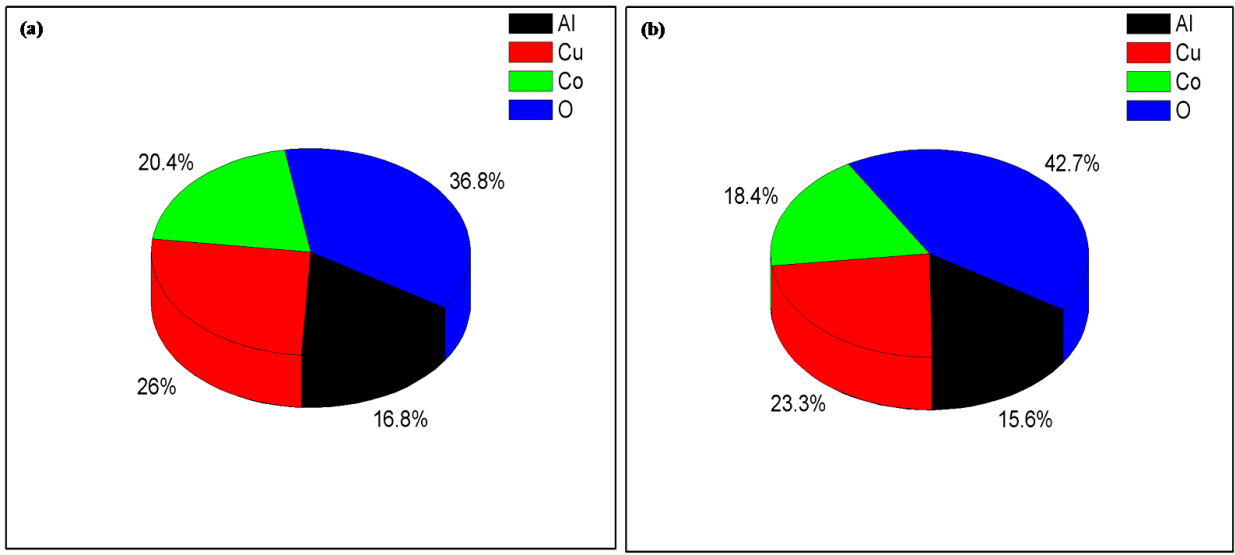

*Fig.9 Element distribution on the surface of 8h leached Al-Cu-Co dQC of (a) $Al_{65}Cu_{15}Co_{20}$ (b) $Al_{65}Cu_{20}Co_{15}$ alloys respectively.*

The energy dispersive x-ray (EDX) analysis confirms that the concentration of Al is quite low compared to Co and Cu. The EDX analysis was done from several areas of leached surface of the DQC alloys and a representative graphical image from 8h leached $Al_{65}Cu_{15}Co_{20}$ and $Al_{65}Cu_{20}Co_{15}$ DQC alloys surface have been given in figures 9 (a-b). The EDX analysis suggests a significant removal of Al and evolution of Co and Cu on the surface along with small amount of oxygen. The oxygen observed on surface of the leached DQC alloys suggested that during leaching and cleaning oxidization proceeds in two steps. The first is chemisorption of oxygen and second is the formation of a thin oxide film; however, the growth of this film will saturate after few monolayers [37]. This saturation is clearly of a kinetic nature since once a few monolayers of oxide were formed, the rate of $O_2$ chemisorption and the ionic transport through the protective oxide layer will be extremely low at room temperature. The distribution of different constituent element after leaching was observed by colour mapping. The colour mapping of EDX analysis shows homogeneous distributions of Co and Cu along with oxygen on the 8h leached surface of the both ($Al_{65}Cu_{15}Co_{20}$ and $Al_{65}Cu_{20}Co_{15}$) DQC alloys as shown in figure 10 (a-b) respectively. A phase separation between Co and Cu was observed in $Al_{65}Cu_{20}Co_{15}$ leached alloy as shown in figure 10(a). However, this type effect was not visible in $Al_{65}Cu_{15}Co_{20}$ leached alloy (figure 10 (b)). It may be noticed that no other contamination e.g. through like Na has been detected.

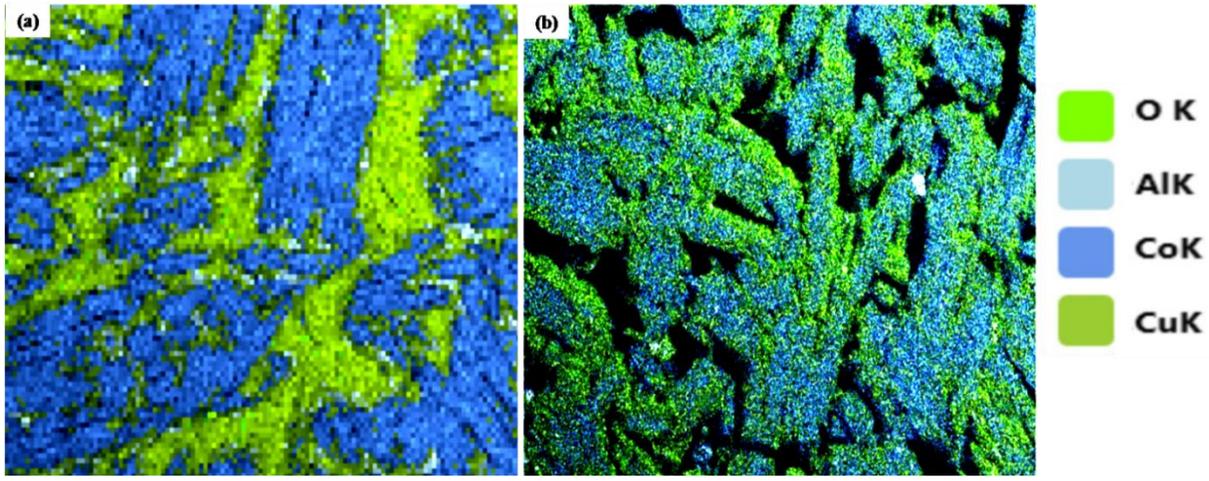

*Fig.10 Constituents elemental distribution on the 8h leached surface (a) $Al_{65}Cu_{15}Co_{20}$ (b) $Al_{65}Cu_{20}Co_{15}$ alloys respectively.*

## 4. Conclusion

The single phase DQC alloy with $Al_{65}Cu_{15}Co_{20}$ and $Al_{65}Cu_{20}Co_{15}$ compositions were chemically leached with 10 mole NaOH alkaline solution for 0.5, 1, 2, 4 and 8 h at room temperature. The formation of homogeneous Cu, Co and their oxide nano particles have been observed in the layer that is generated by the leaching treatment. The microstructure of the leached layer is predominantly controlled by the dissolution rate of Al.

## Acknowledgements

The authors would like to thank, Prof. R.S. Tiwari, Prof. N.K. Mukhopadhyay, Prof. M.A. Shaz and Dr. Hem Raj Sharma for their encouragement and helpful discussion.